\documentstyle[prl,aps,preprint,psfig]{revtex}

\begin{document}
\draft
\preprint{TU-558, RCNS-98-20}
\title{
Hadronic Axion Model in Gauge-Mediated Supersymmetry Breaking
and Cosmology of Saxion
}
\author{T. Asaka}
\address{Institute for Cosmic Ray Research, University of Tokyo,
  Tanashi 188-8502, Japan}
\author{Masahiro Yamaguchi}
\address{Department of Physics,  Tohoku University, Sendai 980-8579, Japan}
\date{November 1998}

\maketitle
\begin{abstract}
  Recently we have proposed a simple hadronic axion model within 
  gauge-mediated supersymmetry breaking.  In this paper we discuss
  various cosmological consequences of the model in great detail. 
  A particular attention is paid to a saxion,
  a scalar partner of an axion, 
  which is produced as a coherent oscillation in the early universe.
  We show that our model is cosmologically viable,
  if the reheating temperature of inflation is sufficiently low.
  We also discuss the late decay of the saxion
  which gives a preferable power spectrum of the density fluctuation 
  in the standard cold dark matter model 
  when compared with the observation.
\end{abstract}

\clearpage

\section{Introduction}
%
The most attractive candidate for the solution of the strong CP
problem is  the Peccei-Quinn (PQ)
mechanism\cite{PQ}. In this mechanism, there exists an axion which is
the Nambu-Goldstone (NG) boson associated with the global $U(1)_{PQ}$
PQ symmetry breaking.

In the framework of gauge mediated supersymmetry (SUSY) breaking
theories\cite{GMSB}, we have proposed an interesting
possibility\cite{A-Y} to dynamically generate the PQ symmetry breaking
scale in a so-called hadronic axion model\cite{KSVZ}.  A gauge
singlet PQ multiplet $X$ and colored PQ quark multiplets $Q_P$ and
$\overline{Q}_P$ are introduced with the superpotential
\begin{eqnarray}
        \label{sp-X}
    W = \lambda_{P} X Q_P \overline{Q_P},
\end{eqnarray}
where $\lambda_P$ is a coupling constant.
Here we stress that no mass parameter is introduced.
Because in a supersymmetric limit the  $U(1)_{PQ}$ symmetry is 
enhanced to its complex extension, there appears a flat direction 
$Q_P = \overline{Q}_P = 0$ with $X$ undermined in the same limit.
The effects of the SUSY breaking, however, lift the flat direction
and fix the vacuum expectation value (vev) of the scalar PQ field $X$.
In Ref.\cite{A-Y} we have shown that the balance of the SUSY breaking effects
between the gravity mediation and the gauge mediation 
stabilizes $X$ and gives the nonzero vev $<X> = f_{PQ}$ approximately as
\begin{eqnarray}
    \label{f_PQ}
    f_{PQ} \simeq \frac{ f^2 }{ m_\sigma }, \label{eq:PQscale}
\end{eqnarray}
where $m_\sigma$ denotes the  mass of a saxion field, the real part 
of the scalar component of
$X$. We estimate
$m_s = \xi m_{3/2}$ with a parameter $\xi$ of order unity.
And $f$ is a mass scale fixed by the messenger sector
of the gauge meditation mechanism.%
\footnote{
  $f$ is roughly represented as $f^2 \sim \left( \frac{\alpha_s}{4 \pi}
  \right) \langle F \rangle$, 
  where $\langle F \rangle$ denotes the F-component vev of the messenger
  singlet field and $\alpha_s$ is the QCD coupling.
}
A current scalar lepton mass limit suggests that
$f \gtrsim 10^4$ GeV.
In the following analysis, we take the minimum value $f=10^4$ GeV
for a  representative value.

The axion arises when the $X$ field develops the non-vanishing vev. 
The decay constant of the axion (the PQ scale) $f_{PQ}$ is constrained
by various astrophysical and cosmological considerations. 
In particular the cooling of the SN 1987A puts a lower bound as
$f_{PQ} \gtrsim 10^{9}$ GeV \cite{SN1987A}.
On the other hand, the upper bound on $f_{PQ}$ comes 
from the overclosure limit of the axion abundance.
The relic abundance of the axion due to the vacuum misalignment is
given by\cite{Turner,EU}
\begin{eqnarray}
    \Omega_a h^2 \sim 0.2 \theta^2
    \left( \frac{ f_{PQ} }{ 10^{12}~\mbox{GeV} } \right)^{1.18},
\end{eqnarray}
with $\theta$ being an initial misalignment angle ($|\theta| < \pi$)
and $h$ the present Hubble constant in units of 100 km/sec/Mpc.
Thus this leads to  $f_{PQ} \lesssim 10^{12}$ GeV.
Here it has been assumed that 
no entropy is produced after the QCD phase transition.
If extra entropy production mechanism with a reheat temperature $T_R$ 
is operated after the transition,
the estimate of the relic abundance changes as \cite{LSSS,K-M-Y}
\begin{eqnarray}
    \Omega_a h^2 \sim 40 \theta^2
    \left( \frac{ T_R }{ 10~\mbox{MeV}} \right)
    \left( \frac{ f_{PQ} }{ 10^{16} ~\mbox{GeV} } \right)^2.
\end{eqnarray}
Then the upper bound on the PQ scale becomes
$f_{PQ} \lesssim 10^{16}$ GeV for $T_R \simeq 10$ MeV.
Note that the big bang nucleosynthesis implies that the reheating 
temperature should be higher than  about 10 MeV.

It follows from Eq. (\ref{eq:PQscale}) that, in the model we are considering,
the allowed region for the PQ scale discussed above
\begin{eqnarray}
   10^9 ~\mbox{GeV} ~\lesssim f_{PQ} 
        \lesssim  10^{12} (10^{16})~\mbox{GeV}
\end{eqnarray} 
can be translated to the allowed region for the saxion mass
\begin{eqnarray}
  100 ~\mbox{MeV} \gtrsim m_\sigma \simeq m_{3/2} \gtrsim 
  100 (10^{-2})~\mbox{keV}.
\end{eqnarray}
Remarkably many models of the GMSB predict a gravitino mass in the
above range.  Thus the model gives a very simple description of the PQ
breaking mechanism solely governed by the physics of the SUSY
breaking.  

In the model, however, 
the saxion has a very light mass which is comparable to 
the gravitino mass and its lifetime may be long.
Furthermore the saxion might be produced as a coherent oscillation too much.
Therefore, 
in this paper we will consider the cosmology of the model,
especially the saxion cosmology, in detail.

The paper is organized as follows.
In the subsequent section, we will discuss cosmological evolution
of the saxion and estimate its relic abundance.
In Section II, we will consider cosmological constraints on 
the saxion and show that our model survives the constraints
if the reheating temperature of the inflation is sufficiently low.
Then in Section IV, we will point out the possibility that the saxion 
plays a role of a late decaying particle,
giving a better fit of the density perturbation in the cold dark matter
scenario.
Final section is devoted to conclusions.
%
\section{Cosmological Evolution of Saxion}
%
Let us begin by discussing the cosmological evolution of the saxion
particle in an inflationary universe, and estimate its relic abundance.
Since the saxion has interaction suppressed by the energy scale $f_{PQ}$,
it decouples from the thermal bath of the universe 
when the cosmic temperature becomes about\cite{RTW}
\begin{eqnarray}
    \label{Tdec}
    T_{dec} \sim 10^{9}~\mbox{GeV}~
    \left( \frac{ f_{PQ} }{ 10^{11} ~\mbox{GeV} } \right)^2
    \sim 10^{9}~\mbox{GeV}~
    \left( \frac{ f }{10^4~\mbox{GeV} } \right)^4
    \left( \frac{ 1~\mbox{MeV} }{ m_\sigma } \right)^2.
\end{eqnarray}
If the temperature achieved after the primordial inflation
was higher than $T_{dec}$, the saxion could be thermalized and
the ratio of the energy density of the saxion $\rho_\sigma$ 
to the entropy density $s$ is estimated as
\begin{eqnarray}
    \label{thermal-1}
    \frac{ \rho_\sigma }{ s }
    \sim 10^{-6} ~\mbox{GeV}~
    \left( \frac{ m_\sigma }{ 1~\mbox{MeV} } \right).
\end{eqnarray}
Even if the saxion was not thermalized,
the saxion could be produced by  scattering processes
right after the reheating era.
In this case the saxion abundance ($\rho_\sigma / s$) is given by
\begin{eqnarray}
    \label{thermal-2}
    \frac{ \rho_\sigma }{ s }
    &\sim& 10^{-6} ~\mbox{GeV}~
    \left( \frac{ m_\sigma }{ 1~\mbox{MeV} } \right)
    \left( \frac{ T_{R} }{ T_{dec} } \right),
    \nonumber \\
    &\sim& 10^{-17} ~\mbox{GeV}~
    \left( \frac{ m_\sigma }{ 1~\mbox{MeV} } \right)^{3}
    \left( \frac{ f }{10^4~\mbox{GeV} } \right)^{-4}
    \left( \frac{ T_R }{ 10~\mbox{MeV} } \right),
\end{eqnarray}
where $T_R$ is the reheating temperature of the inflation 
($T_R \gtrsim 10$ MeV)
and it is highly model dependent\cite{Linde}.
In the gauge-mediated SUSY breaking models, 
since the longitudinal component of the light gravitino 
has interaction much stronger than the gravitational one, 
it would be abundant and overclose the universe 
unless the mass is lower than about 1 keV. 
To cure this, 
the reheating temperature $T_R$ must be sufficiently low\cite{M-M-Y,G-M-M}%
\footnote{
If one assumes that the universe experienced the
late-time entropy production with a sufficiently low reheating temperature,
the gravitino is sufficiently diluted
and we have no upper bound on the reheating temperature of the primordial
inflation $T_R$ 
like Eq.({\protect \ref{tr-gra}}).
}
\begin{eqnarray}
    \label{tr-gra}
    T_R \lesssim
    10^6 ~\mbox{GeV}~
    \left( \frac{300~\mbox{GeV}}{m_{\widetilde{g}}} \right)^2
    \left( \frac{ m_{3/2} }{ 1~\mbox{MeV} } \right).
\end{eqnarray}
With such a low $T_R$, 
the relic abundance of the saxion produced by the scattering
in the thermal bath is highly suppressed.

On the other hand, the saxion is also produced as a coherent oscillation.
The scalar potential in general receives corrections at an early universe, e.g.
during the primordial inflation. Thus 
the saxion does not sit at the true minimum $\sigma=f_{PQ}$ 
but is generically displaced at $\sigma = \sigma_{I}$ 
after the inflation.
Knowledge of  the initial displacement $\sigma_I$ is crucial 
to estimate the abundance of the oscillating saxion, but it depends on
the details of supergravity Lagrangian as well as inflation models.%
\footnote{
  $\sigma_I \sim 0$ or $\sigma_I \sim M_G$ may be naturally expected
  by the supergravity effect.
  }
Thus in the following, 
we regard $\sigma_I$ as an arbitrary parameter which is not larger than 
the Planck scale 
($M_G = 2.4 \times 10^{18}$ GeV) and consider two distinct cases.
%
\subsection{Case 1: Saxion trapped at the origin}
%
Let us first consider the case that the saxion field sits 
around the origin ($\sigma_I \simeq 0$) just after the inflation.%
\footnote{
Thus the $U(1)_{PQ}$ symmetry is recovered.
In this case it should be noted that,
since the effective mass which the saxion feels during the 
primordial inflation 
is generally comparable to the Hubble parameter during the inflation,
the quantum fluctuations for the $X$ field 
are strongly suppressed{\protect \cite{Enquvist-NG-Olive}}.
For special K\"{a}hler potential, a massless mode appears in the $X$ field
and a large quantum fluctuation is expected.
This would lead to too much axion isocurvature fluctuation.
To suppress it, the mass density of the axion 
or the Hubble parameter during the inflation should be small.
We thank M. Kawasaki and T. Yanagida for pointing out this problem.
}
In this case the PQ quark multiplets ($Q_P$ and $\overline{Q_P}$)
are in the thermal equilibrium through the rapid QCD interaction, which 
generates an  additional mass term $\sim T^2 |X|^2$ 
to the saxion potential as a finite temperature effect.
Thus the saxion is trapped at the origin till the cosmic temperature 
becomes below a critical temperature $T_C$.
We expect that $T_C$ $\sim 100$ GeV,
since the saxion obtains a negative soft SUSY breaking mass
squared at the origin due to the renormalization group effect 
of the Yukawa interaction Eq.(\ref{sp-X})
and its absolute value is expected to be of the order of 
the electroweak scale $\sim 100$ GeV.

Therefore the saxion starts to roll down toward its true minimum 
at $T=T_C$ and oscillates around it with the initial amplitude 
$\sigma_0 \simeq f_{PQ}$.%
\footnote{
  We are free from the domain wall problem in our model,
  since $N_{DW} =1$.
}
The ratio of the energy density of this oscillation to
the entropy density is estimated as
\begin{eqnarray}
    \label{Rs0}
    \frac{ \rho_\sigma }{ s } 
    \simeq 
    \frac{ {\displaystyle
           \frac{ 1 }{ 2 } m_\sigma^2 f_{PQ}^2
           \left( \frac{\sigma_0}{f_{PQ}} \right)^2 } }
         { \frac{ 2 \pi^2 }{45} g_\ast T_C^3 },
\end{eqnarray}
where $g_\ast$ counts the degrees of freedom of the relativistic
particles ($g_\ast \sim 100 $ at $T=T_C$).
Here it should be noted that this ratio takes a constant value 
till the saxion decays if no extra entropy is produced,
since $\rho_\sigma$ and $s$ are both diluted 
by $\rho_\sigma \propto s \propto R^{-3}$ 
($R$: the scale factor of the universe) as the universe expands.
In the model we are discussing, the PQ scale $f_{PQ}$ is given by
Eq.(\ref{f_PQ})  and thus
the abundance of the saxion oscillation reads 
\begin{eqnarray}
    \label{Rs-01}
    \frac{ \rho_\sigma }{ s }
    &\simeq &
    1.1 \times 10^{-2} \frac{ f^4 }{ T_C^3 }
    \left( \frac{\sigma_0}{f_{PQ}} \right)^2,
    \nonumber \\
    &\simeq &
    1.1 \times 10^{8} ~\mbox{GeV}~
    \left( \frac{ f }{10^4~\mbox{GeV}} \right)^4
    \left( \frac{ T_C }{ 100~\mbox{GeV} } \right)^{-3}
    \left( \frac{\sigma_0}{f_{PQ}} \right)^2.
\end{eqnarray}
Note that it is independent of the saxion mass.
Here we have assumed that the reheating process of the 
primordial inflation completed before the saxion 
oscillation starts, i.e., $T_R > T_C$.
On the other hand, for the case  $T_R < T_C$,
we expect a dilution of  the saxion energy density.
The saxion energy density at $T = T_R$ is given by
\begin{eqnarray}
    \rho_\sigma (T_R) \simeq
    \rho_\sigma (T_C) 
    \left( \frac{ R(T_C) }{ R(T_R) } \right)^3
    \simeq
    \frac{1}{2} m_\sigma^2 f_{PQ}^2
    \left( \frac{ \sigma_0 }{ f_{PQ} } \right)^2
    \left( \frac{ R(T_C) }{ R(T_R) } \right)^3.
\end{eqnarray}
From the fact that the radiation energy density is proportional to 
$R^{-3/2}$ from the end of the inflation till $T = T_R$,  it follows that
the ratio of the scale factor ($R(T_C)/R(T_R)$) becomes
\begin{eqnarray}
    \left( \frac{ R(T_C) }{ R(T_R) } \right)^{3/2}
    =
    \frac{ g_\ast(T_R) T_R^4 }{ g_\ast(T_C) T_C^4 }.
\end{eqnarray}
Thus the saxion abundance at $T = T_R$ is given by
\begin{eqnarray}
    \label{Rs-02}
    \frac{ \rho_\sigma }{ s }
    &\simeq&
    1.2 \times 10^{-3}
    m_\sigma^2 f_{PQ}^2 \frac{ T_R^5 }{ T_C^8 }
    \left( \frac{ \sigma_0 }{ f_{PQ} } \right)^2,
    \nonumber \\
    &\simeq&
    1.2 \times 10^{-13} ~\mbox{GeV}~ 
    \left( \frac{ f }{ 10^4 ~\mbox{GeV} } \right)^4
    \left( \frac{ T_C }{ 100 ~\mbox{GeV} } \right)^{-8}
    \left( \frac{ T_R }{ 10~\mbox{MeV} } \right)^5
    \left( \frac{ \sigma_0 }{ f_{PQ} } \right)^2.
\end{eqnarray}
Here note that 
the reheating temperature $T_R$ should be higher than about 10 MeV
in order to maintain the success of the big bang nucleosynthesis.
Comparing it with Eq.(\ref{Rs-01}), we find that 
the relic abundance is diluted significantly
by the primordial inflation with a very low $T_R$.

The above discussion does also hold even for the case
where the saxion is displaced from the origin 
$\sigma_I$ by a small amount.
This is because the PQ multiplets will be thermalized and generate the 
potential which traps the saxion
 at the origin,
as far as their effective mass ($\sim \lambda_P \sigma_I$) 
is smaller than the maximum temperature achieved after 
the primordial inflation.%
\footnote{
The maximum temperature $T_{MAX}$ achieved after 
the primordial inflation is given by\cite{EU} as
$T_{MAX} \simeq 0.3 T_R^{1/2} V_I^{1/8}$,
where $V_I$ is the vacuum energy of the inflation.
}
%
\subsection{Case 2: Saxion displaced far from the origin}
%
If the displacement $\sigma_I$ is large enough,
the cosmological evolution of the saxion is completely different.
Here, we consider only the case that 
$( M_G \gtrsim )\sigma_I \gtrsim f_{PQ}$.%
\footnote{
Thus the $U(1)_{PQ}$ symmetry is broken and  a massless
mode, i.e. the axion, appears. 
See the footnote 4 for an argument on the density fluctuation 
of the axion field.
}
When the Hubble parameter of the universe ($H$) becomes $H \simeq
m_\sigma$,
the saxion starts to obey the coherent oscillation.%
\footnote{
Note that for the region $\sigma = \sigma_I \gtrsim f_{PQ}$ 
the potential is governed by the contribution from the gravity
mediation effect, and the effective mass of the saxion
is almost the  same as the gravitino mass, i.e., the saxion mass 
($m_{3/2} \simeq m_\sigma$).
}
The temperature at this time $T_\sigma$ is estimated as
\begin{eqnarray}
    \label{T_phi}
    T_\sigma 
    &\simeq& 0.46
    \sqrt{ m_\sigma  M_G },
    \nonumber\\
    &\simeq& 
    2.3 \times 10^7 ~\mbox{GeV} 
    \left( \frac{ m_\sigma }{ 1 ~\mbox{MeV} } \right)^{1/2}.
\label{eq:Tsigma}
\end{eqnarray}
Using the initial amplitude of the oscillation 
$\sigma_0 = \sigma_I$ ($\gtrsim f_{PQ}$),
the saxion abundance of this oscillation is estimated as
\begin{eqnarray}
    \label{Rs-11}
    \frac{ \rho_\sigma }{ s }
    &\simeq&
     \frac{ {\displaystyle
     \frac{ 1 }{ 2 } m_\sigma^2 M_G^2 
            \left( \frac{ \sigma_0 }{ M_G } \right)^2 } }
          { \frac{ 2 \pi^2 }{ 45 } g_\ast T_\sigma^3 } 
    \nonumber \\
    &\simeq&
    2.8 \times 10^{ 6 } ~\mbox{GeV}~
    \left( \frac{ m_\sigma }{ 1 ~\mbox{MeV} } \right)^{1/2}
    \left( \frac{ \sigma_0 }{ M_G } \right)^2.
\end{eqnarray}
Here we have assumed that the saxion oscillation 
starts after the reheating of the primordial inflation 
had been completed ($T_R > T_{\sigma}$).
However, such a high $T_R$ leads the gravitino problem
for $m_\sigma ~(\simeq m_{3/2}) \gtrsim 1$ keV.
Comparison of Eq.~(\ref{eq:Tsigma}) with the upper bound on 
$T_R$ [Eq.(\ref{tr-gra})] shows that $T_{\sigma}$ is always higher 
than this bound. On the other hand, 
the inflation with $T_R < T_\sigma$ 
dilutes the energy density of the saxion as
\begin{eqnarray}
    \label{Rs-12}
    \frac{ \rho_\sigma }{ s }
    &\simeq &
    \frac{1}{8} T_R 
    \left( \frac{ \sigma_0 }{ M_G } \right)^2
    \nonumber \\
    &\simeq&
    1.3 \times 10^{-3} ~\mbox{GeV}~
    \left( \frac{ T_R }{ 10~\mbox{MeV} } \right)
    \left( \frac{ \sigma_0 }{ M_G } \right)^2.
\end{eqnarray}
This ratio takes its minimum value for 
the lowest reheating temperature $T_R \simeq 10$ MeV
and the smallest initial amplitude $\sigma_0 \simeq f_{PQ}$.

To end this section we estimate the saxion lifetime.
When the Hubble parameter becomes comparable to its decay
width, the saxion decay occurs.
In the present model, the saxion dominantly decays into two axions
with the rate%
\footnote{
The saxion may decay into two photons with a very small 
branching ratio.
We will discuss this rare process in the next section.
}
\begin{eqnarray}
    \label{gam-s} 
    \Gamma (\sigma \rightarrow a a )
    =
    \frac{ 1 }{ 64 \pi }
    \frac{ m_\sigma^3}{ f_{PQ}^2 },
\end{eqnarray}
and the lifetime is given as
\begin{eqnarray}
    \label{tau-s}
    \tau_\sigma \simeq 
    1.3 \times 10^{9} ~\mbox{sec}~
    \left( \frac{ f }{ 10^4~\mbox{GeV} } \right)^4
    \left( \frac{ m_\sigma }{ 1 ~\mbox{MeV} } \right)^{-5}.
\end{eqnarray}
Therefore the saxion becomes stable within the age of the universe
($\sim 10^{17}$ sec) if its mass is smaller than about 10 keV. 
Note that the value $\tau_\sigma$ varies very widely
because of the strong dependence on the saxion mass.

%
\section{Constraints on Cosmic Saxion}
%
In this section we consider the cosmological constraints on the saxion
abundance and examine whether the proposed model is viable or not.
Since the energy density of the saxion is not so suppressed compared to
the radiation and also it may have a long lifetime, we have to
carefully see its cosmological effects.

As described in the previous section, the saxion is produced by 
the scattering at the reheating era 
and also as the coherent oscillation.
Comparing the  both saxion abundances with each other, one finds that
the saxion produced from the thermal bath can be neglected,
if one assumes the primordial inflation 
with a relatively low reheating temperature 
to avoid the gravitino problem 
and too much relic abundance of the saxion oscillation.
Therefore it should be enough to consider only the saxion abundance coming
from the oscillation in the subsequent discussion.
%
\subsection{Constraints on Saxion Abundance}
%
First of all,
the saxion with $\tau_\sigma \gtrsim 1$ sec,
i.e., $m_\sigma \lesssim 100$ MeV,
may upset the big bang nucleosynthesis.
Existence of the extra energy of exotic particle at $T\sim 1$ MeV 
may speed up the expansion of the universe and 
increase the number ratio of neutron to proton, resulting in 
the overproduction of $^4$He.
Roughly speaking, the energy of such a particle should
be smaller than that of one neutrino species.
Thus the saxion abundance should satisfy
\begin{eqnarray}
    \label{BBN-1}
    \frac{ \rho_\sigma }{ s } \lesssim 10^{-4} ~\mbox{GeV}.
\end{eqnarray}

Furthermore, if the saxion lifetime is larger than 
the age of the universe (i.e., $m_\sigma \lesssim 10$ keV), 
the saxion oscillation still exists now.
Not to overclose the universe, 
the saxion abundance should be 
\begin{eqnarray}
    \label{Rcr}
    \frac{ \rho_\sigma }{ s } \lesssim \frac{ \rho_{c} }{ s_0 } 
    = 3.6 \times 10^{-9} ~h^2~\mbox{GeV},
\end{eqnarray}
with the critical density $\rho_c$ 
and the present entropy density $s_0$.

Next, we turn to the rare decay of the saxion.
Since the saxion is light enough in our model,
it can not decay into two gluons 
but  decays into two photons via one-loop diagrams 
if the PQ quark multiplets have QED charges.
Thus we expect that the branching ratio of this radiative decay 
is very suppressed as 
$B_\gamma \sim ( \alpha_{em} /(4\pi) )^2 \sim 10^{-7}$.
For example, if 
the PQ quark fields $Q_P$ and $\overline{Q}_P$ 
are in $\bf 5$ and $\bf 5^{\ast}$ under the SU(5) GUT gauge group,
the branching ratio of the radiative decay is given as 
$B_\gamma \simeq 5 \times 10^{-6}$.
Even with such a small $B_\gamma$,
extra photons produced at a late time 
are cosmologically dangerous.

First, the saxion decay  is constrained from the diffuse x-ray
background spectrum.
When the saxion has a lifetime longer than the cosmic time
of the recombination ($\tau_\sigma \gtrsim 10^{12}$ sec),
the produced photon directly contributes to the spectrum,
and then a very stringent upper bound on 
$B_\gamma \times (\rho_\sigma/s)$
comes from the present observation of the spectrum.
(The details are found, for example, in Ref.\cite{K-Y}.)

If the saxion lifetime is in between  $10^6$ sec and $10^{12}$ sec, 
extra radiation energy produced by the saxion decay
may alter the cosmic microwave background spectrum from the blackbody one.
The observation by the COBE satellite\cite{COBE-CMBR} 
gives the following upper bounds on the saxion abundances as
\begin{eqnarray}
    \label{CMBR-1}
    &&B_\gamma \frac{ \rho_\sigma }{ s }
        \lesssim 2.5 \times 10^{-5} ~ T_D
    ~~~~\mbox{for}~
    10^6~\mbox{sec} \lesssim \tau_\sigma \lesssim
    10^{10}~\mbox{sec},\\
    \label{CMBR-2}
    &&B_\gamma \frac{ \rho_\sigma }{ s }
        \lesssim 2.3 \times 10^{-5} ~  T_D
    ~~~~\mbox{for}~
    10^{10}~\mbox{sec} \lesssim \tau_\sigma \lesssim
    10^{12}~\mbox{sec},
\end{eqnarray}
where $T_D$ is the cosmic temperature at the saxion decay
and estimated as
\begin{equation}
        \label{TD}
        T_D = 0.60 ~\frac{ \rho_\sigma }{ s }
        \left[ 
            \frac{ M_G }{
             {\displaystyle
             \left( \frac{ \rho_\sigma }{ s } \right)^2 \tau_\sigma }
             }  
        \right]^{2/3}.
\end{equation}

These constraints on the saxion abundance are 
summarized in Fig.\ref{fig:RsBr5d-6} 
for the case $B_\gamma = 5 \times 10^{-6}$.
Note that the constraints from the x-ray background spectrum
and the CMBR become weaker for the suppressed $B_\gamma$.
We can now compare these cosmological constraints 
with the saxion abundance estimated in the previous section.
%
\subsection{Case 1: Saxion trapped at the origin}
%
First, we consider the case that the saxion is thermally trapped
at the origin after the primordial inflation.
For the case that the reheating temperature of the inflation is 
higher than the critical temperature ($T_R > T_C \sim 100$ GeV),
the saxion abundance [Eq(\ref{Rs-01})] significantly exceeds the 
upper bound Eq.(\ref{BBN-1}) from the big bang nucleosynthesis.
To avoid this the saxion should decay before it,
however the lifetime of the saxion is longer than 1 sec
if $m_\sigma \lesssim 100$ MeV.
Even for the case $m_\sigma \simeq 100$ MeV,
the energy of the decay produced axion easily exceeds
the bound Eq.(\ref{BBN-1}).
Therefore, in this case, we conclude that $T_R < T_C$.

\begin{figure}
    \centerline{\psfig{figure=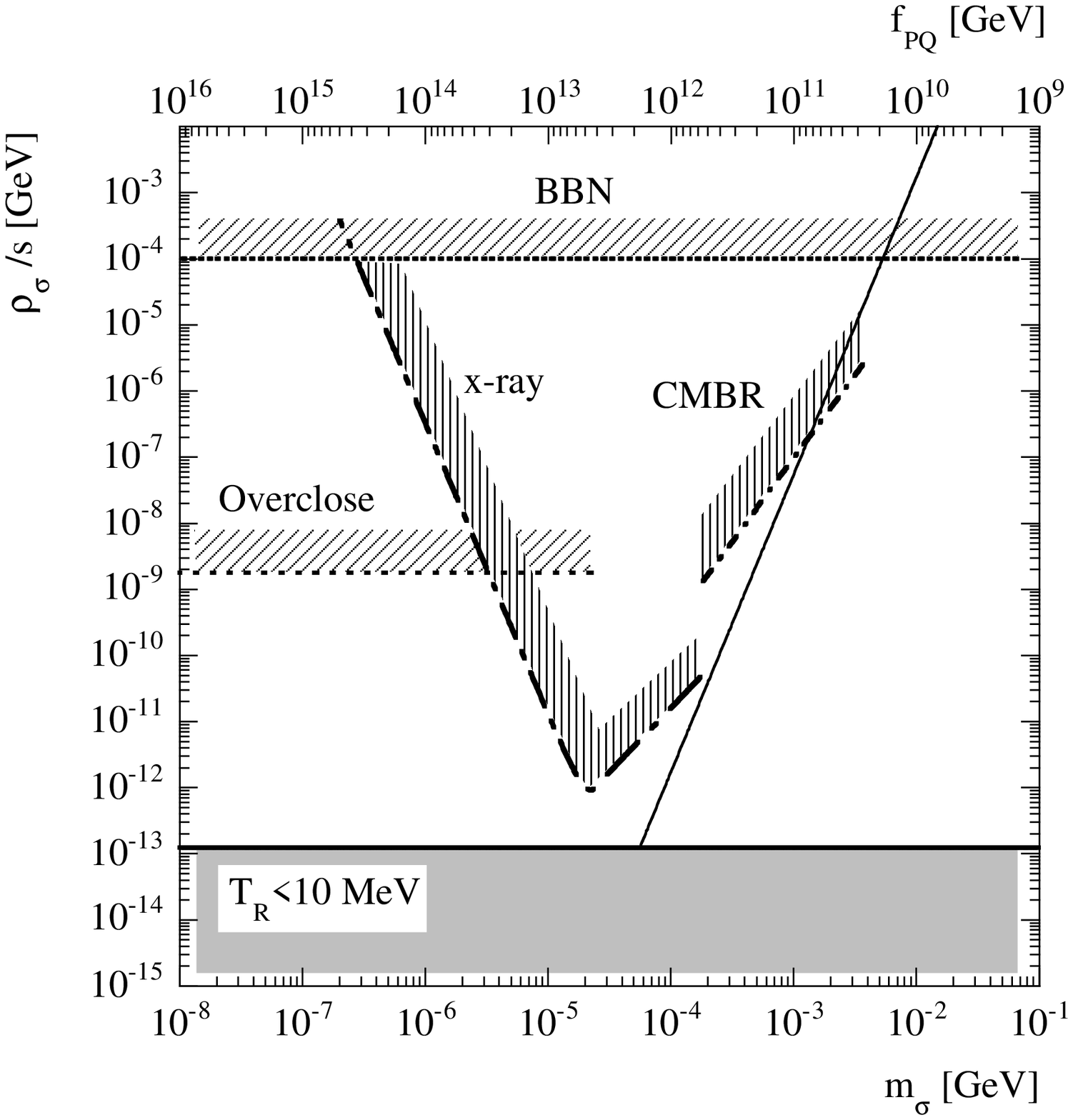,width=15cm}}
    \vspace{0.5cm}
\caption{The lower bound of the saxion abundance [Eq.(\protect \ref{Rs-02})]
for the case $T_C =100$ GeV and $f=10^{4}$ GeV (the thick solid line).
We take the value of the radiative branching
ratio $B_\gamma = 5 \times 10^{-6}$.
The upper bounds from the various cosmological constraints 
are also shown.
For the saxion mass larger than about 50 keV,
the predicted saxion abundance in the new inflation model
{\protect \cite{I-K-Y}}  is shown with the thin solid line.}
    \label{fig:RsBr5d-6}
\end{figure}

The saxion abundance [Eq.(\ref{Rs-02})] for $T_R < T_C$
takes its minimum value at the lowest $T_R = 10$ MeV, since the dilution
of the inflation becomes most effective.
In Fig.\ref{fig:RsBr5d-6}
we show this lower limit of the saxion abundance with various 
cosmological constraints.
Note that now for the case $T_R =10$ MeV,
the PQ scale as high as $10^{16}$ GeV is allowed in the consideration 
of the relic abundance of the axion oscillation. 
It is found that all the region for the saxion mass
is cosmologically allowed if we take the sufficiently low 
reheating temperature $T_R \simeq 10$ MeV.
Furthermore  the stable saxion with mass $10$ eV $\lesssim m_\sigma 
\lesssim$ 10 keV will constitute a dark matter of our universe,
since the saxion could achieve $\Omega_\sigma =
(\rho_\sigma/\rho_c)_0 = 1$ without conflicting with the cosmic x-ray
background observation.

In the above result an extremely low 
reheating temperature is crucial to dilute the saxion 
(and also the gravitino).
The new inflation model\cite{New-Inflation} gives such a situation.
For example, Ref.\cite{I-K-Y} gives a new inflation model
which solves the initial condition problem,
where the reheating temperature is given by
\begin{eqnarray}
    \label{TR-IKY}
    T_R \sim m_{3/2}^{9/10} M_G^{1/10},
\end{eqnarray}
where the gravitino mass should be larger than about 50 keV 
since $T_R \gtrsim$ 10 MeV.
With this reheating temperature the saxion abundance (\ref{Rs-02})
is also shown in Fig.\ref{fig:RsBr5d-6}.
It can be seen that with this new inflation model\cite{I-K-Y}
the saxion with $m_\sigma \simeq$ 50 keV-1 MeV survives 
the various cosmological constraints.

%
\subsection{Case 2: Saxion displaced far from the origin}
\begin{figure}
    \centerline{\psfig{figure=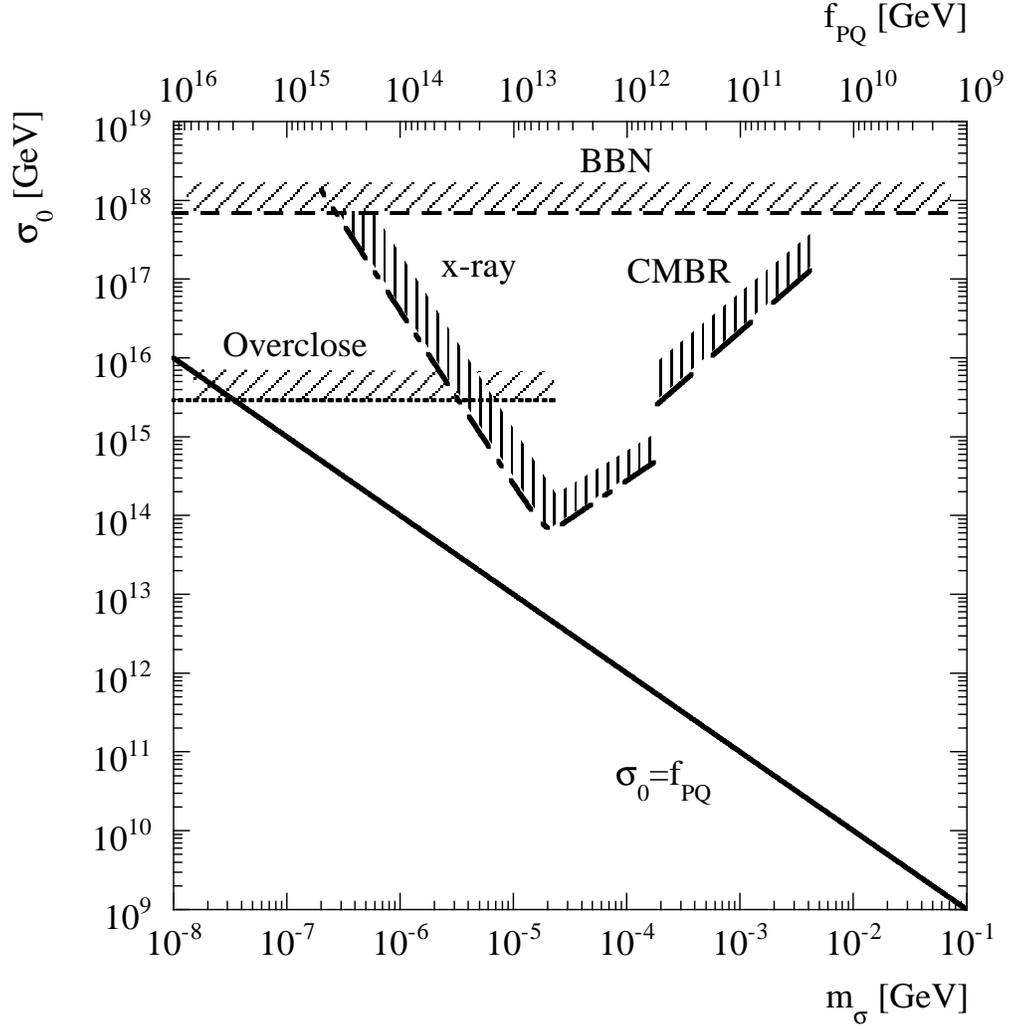,width=15cm}}
    \vspace{0.5cm}
\caption{
Various cosmological upper bounds on the initial amplitude
of the saxion oscillation 
[Eq.({\protect \ref{Rs-12}})] for the case $T_R =$ 10 MeV 
($\ll T_{\sigma}$).
We take $f=10^4$ GeV and $B_\gamma = 5\times10^{-6}$.
We also show the initial amplitude $\sigma_0 = f_{PQ}$
by the solid line.
}
    \label{fig:ConstS0}
\end{figure}

%
Next we turn to the case that the saxion is displaced far from 
the origin after the primordial inflation.
In order to avoid the overproduction of the gravitino 
we assume  the reheating temperature of the inflation to
be lower than $T_\sigma$ (even for the case that 
$m_{3/2} \simeq m_\sigma \lesssim $ 1 keV).
The saxion abundance for $T_R \lesssim T_\sigma$
is given as Eq.(\ref{Rs-12}) and 
takes its minimum value when  the minimum $T_R$ = 10 MeV.
In Fig.\ref{fig:ConstS0}
we show the cosmological constraints on the initial amplitude
$\sigma_0$ in the saxion abundance Eq.(\ref{Rs-12}) with 
$T_R$ = 10 MeV.
Note that in this case the upper bound on the PQ scale 
is $f_{PQ} \lesssim 10^{16}$ GeV.
We find that most of the saxion mass (or the PQ scale)
is cosmologically allowed if we take $\sigma_0 \sim f_{PQ}$.
And when the saxion has a heavier mass $m_\sigma \simeq 10$--100 MeV,
the initial amplitude can take a value 
from $\sigma_0 \sim f_{PQ}$ to $\sigma_0 \sim M_G$.
Similar to the previous case, 
the saxion with mass $10$ eV $\lesssim m_\sigma \lesssim$ 10 keV
could be a dark matter of our universe.

To summarize, the hadronic axion model we proposed is cosmologically
viable in the both cases as far as the reheating temperature of the
inflation is low enough.

%
%
To end this section, we emphasize that 
it has been assumed that  no extra entropy 
production takes place other than the primordial inflation.
If such a late-time entropy production like
a thermal inflation\cite{L-S} or 
an oscillating inflation\cite{G-M-M,Moroi,OI} occurs,
tremendous entropy is released at a late-time in the history
of the universe (but at least before the big bang nucleosynthesis)
and the saxion abundance is diluted by about
$10^{10}$--$10^{20}$ and  cosmological constraints on it
are extensively relaxed. Thus 
our model becomes cosmologically viable even with the 
primordial inflation model with a high reheating temperature.

%
\section{Late Decaying Saxion and Structure of the Universe}
%
As discussed in the previous section, 
the cosmic energy of the saxion oscillation 
is abundant compared to the radiation
and it may decay at very late time of the universe.
In this section, we would like to argue that 
this feature of the saxion particle
could explain well the present observed power spectrum of the 
cosmic structure with some cold dark matter\cite{C-K},
i.e., the saxion can be a candidate for a late decaying 
massive particle (LDP)\cite{LDP1,LDP2}.

It is well known that
the predicted spectrum of the standard cold dark matter
(CDM) with $\Omega_0 \simeq 1$ and $h \simeq 0.7$
could not fit well all the observational data simultaneously.
The cosmic background explorer (COBE) 
observed the anisotropy of the temperature of the cosmic 
microwave background radiation (CMBR)\cite{DMR} 
and gave the normalization of the spectrum 
at a large scale ($\lambda \sim 10^3 h^{-1}$Mpc).
However, this normalization predicts
too much fluctuation at a small scale 
($\lambda \sim 10 h^{-1}$Mpc) than galaxy distribution observations%
\cite{P-D}.
The late decaying massive particle (LDP), 
once dominating the energy of the universe before its decay, 
can delay the epoch of the matter-radiation equality compared 
to the standard CDM model,
which results in the preferred spectrum\cite{LDP1,LDP2}.

\begin{figure}
    \centerline{\psfig{figure=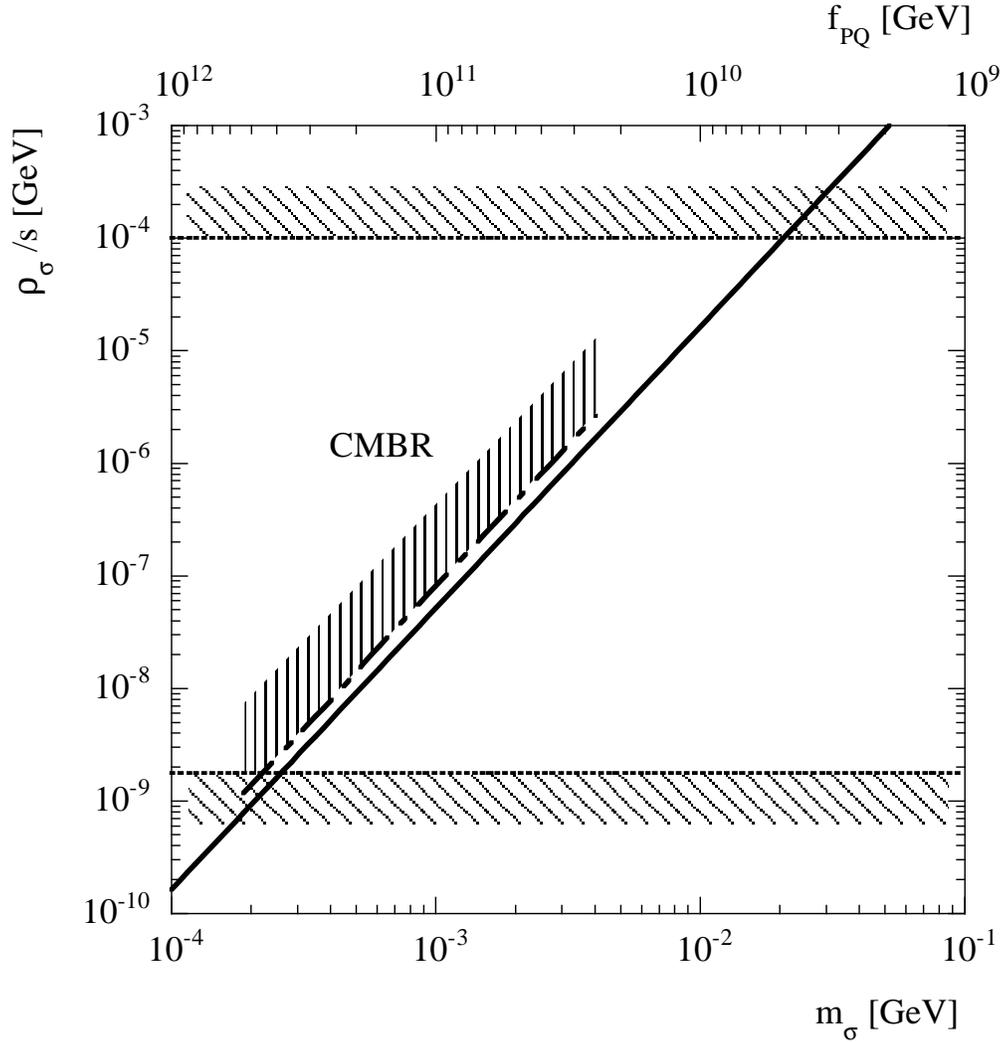,width=15cm}}
    \vspace{0.5cm}
\caption{
The abundance of the saxion which is required to become the LDP
(the solid line).
Here we take $f=10^4$ GeV, $\Omega_0=1$ and $h=0.7$.
The upper bound from the big bang nucleosynthesis and
the lower bound from the saxion domination are denoted by
the dotted lines with shadows.
The dot-dashed lines represent the upper bounds of the saxion
abundance from the CMBR for the cases that
the radiative branching ratio $B_\gamma = 5\times10^{-6}$.
}
    \label{fig:RsLDP}
\end{figure}

Ref.\cite{LDP2} argued that
the saxion can be accounted to be the LDP,
if the saxion energy density satisfies 
\begin{eqnarray}
    \label{cond-LDP1}
        \frac{ \rho_\sigma }{ s }
        \gtrsim \Omega_0 \times \frac{ \rho_c }{ s_0 },
\end{eqnarray}
to cause the saxion domination era, and
\begin{eqnarray}
        \label{cond-LDP2}
        \tau_\sigma \left( \frac{ \rho_\sigma }{ s } \right)^2
        \simeq x M_G,
\end{eqnarray} 
to explain the observed power spectrum.
Here $x$ is defined as
\begin{eqnarray}
        x &=& 0.23 \left[ 
        \left( \frac{ \Omega_0 h }{ 0.3 } \right)^2 - 1 
        \right]^{3/2}.
\end{eqnarray}
In the following we will take $x = 2.2$, which corresponds to 
the case that $\Omega_0 = 1$ and $h=0.7$.
Note that the saxion LDP also survives the big bang nucleosynthesis
constraint (\ref{BBN-1}).
In Fig.\ref{fig:RsLDP} we show the saxion abundance 
required to become the LDP 
by using its lifetime [Eq.(\ref{tau-s})].
In the same figure, we also show
various cosmological constraints.
We find that the saxion whose mass is 
\begin{eqnarray}
  \label{ms-LDP}
  10 \mbox{MeV} \gtrsim m_\sigma \gtrsim 100 \mbox{keV}
\end{eqnarray}
can be the LDP.
However, such a saxion is severely constrained 
by the CMBR constraints [Eq.(\ref{CMBR-1})].%
\footnote{
Note that the saxion with mass $\sim 10$ MeV 
is free from the CMBR constraint since the lifetime 
is short.
}
When the saxion becomes the LDP, the CMBR constraints
give the upper bounds on the radiative branching ratio as
\begin{eqnarray}
    && B_\gamma 
        \lesssim 8.8 \times 10^{-6},
    ~~~~~~~~~\mbox{for}~~~
    10^6~\mbox{sec} \lesssim \tau_\sigma \lesssim
    10^{10}~\mbox{sec},\\
    && B_\gamma 
        \lesssim 8.2 \times 10^{-6},
    ~~~~~~~~~\mbox{for}~~~
    10^{10}~\mbox{sec} \lesssim \tau_\sigma \lesssim
    10^{12}~\mbox{sec}.
\end{eqnarray}
Therefore $B_\gamma = 5\times 10^{-6}$ for the simple case 
that the PQ quark multiplets $Q_P$ and 
${\overline Q}_P$ are $\bf 5$ and $\bf 5^{\ast}$ 
of the SU(5) gauge group is marginally allowed.
However, $B_\gamma$ depends very much on their representation.
Here we claim that the observational bound of the CMBR at present
is comparable to the prediction of the late decaying particle.
In other words the late decaying saxion scenario presented here 
will be tested by  more precise experiments of the CMBR
in near future (e.g. the experiments by the PLANCK and MAP 
satellites).
%
\subsection{Axion CDM and Saxion LDP Scenario}
%
We will explain that
the saxion abundance required for the LDP 
[Eqs.(\ref{cond-LDP1}) and (\ref{cond-LDP2})]
can be naturally offered in the present model.
First, we consider the case that the saxion is trapped at 
the origin due to the thermal effects after the primordial inflation.
Using the relic abundance of the saxion Eq.(\ref{Rs-02}),
the condition for the  LDP [Eq.(\ref{cond-LDP2})] leads to
the required reheating temperature of the inflation as%
\footnote{
For example, a new inflation model by Ref.{\protect \cite{I-K-Y}}
with the reheating temperature ({\protect \ref{TR-IKY}})
leads to the late decaying saxion which mass is
\begin{eqnarray}
    m_\sigma \simeq m_{3/2}
    \simeq  0.92 ~\mbox{MeV} ~
    \left( \frac{ \sigma_0 }{ f_{PQ} } \right)^{-1}
    \left( \frac{ T_C }{100 ~\mbox{GeV} } \right)^{4}
    \left( \frac{ f }{10^{4} ~\mbox{GeV} } \right)^{-3},
\end{eqnarray}
which just lies in the required mass region [Eq.(\ref{ms-LDP})].
}
\begin{eqnarray}
    \label{TR-LDS}
    T_R &\simeq& 2.3 ~x^{1/10} 
    \left( \frac{ \sigma_0 }{ f_{PQ} } \right)^{-2/5}
    \frac{ M_G^{1/10} m_\sigma^{1/2} T_C^{8/5} }{ f^{6/5} },
    \nonumber \\
    &\simeq&
    130 ~\mbox{MeV}~
    \left( \frac{ \sigma_0 }{ f_{PQ} } \right)^{-2/5}
    \left( \frac{ m_\sigma }{ 1 ~\mbox{MeV} } \right)^{1/2}
    \left( \frac{ T_C }{ 100 ~\mbox{GeV} } \right)^{8/5}
    \left( \frac{ f }{ 10^4 ~\mbox{GeV} } \right)^{6/5},
\end{eqnarray}
with the saxion mass Eq.(\ref{ms-LDP}).

It is worth noting that 
if we take the critical temperature as $T_C > 100$ GeV,
the reheating of the inflation, which induces the required 
saxion abundance of the LDP,
completes before the QCD phase transition.
In this case 
the axion can constitute the dominant component of the dark matter
of our universe if $f_{PQ} \sim 10^{11}$--$10^{12}$ GeV.
This PQ scale corresponds to the saxion mass
$m_\sigma \sim$ 100 keV--1 MeV.
Therefore, in the considering model, 
if one takes the gravitino mass 
$m_{3/2} \simeq m_\sigma \simeq$ 100 keV--1 MeV,
the axion can be the CDM of our universe 
and moreover the saxion can be the late decaying particle 
to cure the difficulty in the power spectrum of the structure of 
our universe in the axion CDM model.

Even if the saxion is displaced from the origin after the inflation,
a similar argument holds.
Using the saxion abundance for $T_\sigma > T_R$ [Eq.(\ref{Rs-12})],
the saxion which mass in Eq.(\ref{ms-LDP}) becomes the LDP when
\begin{eqnarray}
  \frac{ \sigma_0 }{ f_{PQ} }
  &\simeq& 0.91
  \frac{ m_\sigma^{9/4} M_G^{5/4} }{ T_R f^3 },
  \nonumber \\
  &\simeq& 1.5 \times 10^3
  \left( \frac{ m_\sigma }{ 1~\mbox{MeV} } \right)^{9/4}
  \left( \frac{ T_R }{ 100~\mbox{GeV} } \right)^{-1/2}
  \left( \frac{ f }{ 10^4~\mbox{GeV} } \right)^{-3}.
\end{eqnarray}
Therefore if we take the reheating temperature as 
$T_R \gtrsim \Lambda_{QCD}$,
the gravitino mass $m_{3/2} \simeq m_\sigma $ 100 keV--1MeV
also leads to the axion CDM.
Thus the axion CDM and the saxion LDP scenario 
is achieved with the moderate $T_R$ and $\sigma_0$.

\subsection{Moduli CDM and Saxion LDP Scenario}
So far we have assumed that there exists no extra dilution mechanism
other than the primordial inflation in the history of the universe.
Here we will briefly discuss the case that 
there exists the late time entropy production such
as the thermal inflation\cite{L-S} or 
the oscillating inflation\cite{G-M-M,Moroi,OI}.
The idea of the late time entropy production 
is that a mini inflation takes place at a late 
time of the universe (at least before the big-bang nucleosynthesis)
and produces tremendous entropy at the reheating epoch,
which dilutes the unwanted particles extensively.
If one assumes this dilution mechanism,
the cosmological constraints on the saxion abundance 
become much weaker
even for the saxion abundance [Eq.(\ref{Rs-11})] with 
$\sigma_0 \sim M_G$.
Here we would like to stress that 
the saxion can play the role of the LDP even in this case.

In the universe with the late time entropy production
it is worth noting that the string moduli could be the CDM.%
\footnote{
Cosmology of the string moduli in the gauge-mediated SUSY breaking
is discussed by Ref.{\protect \cite{G-M-M,H-K-Y,A-H-K-Y}}
in the universe with the thermal inflation,
and by Ref.{\protect \cite{G-M-M,Moroi,A-K-Y}}
in the universe with the oscillating inflation.
}
The moduli are scalar fields associated with flat directions
which characterize classical vacua of the string theory and have only
gravitationally
suppressed interaction.
The effects of the SUSY breaking 
give the moduli masses $m_\phi$ comparable to the gravitino 
mass ($m_\phi \simeq m_{3/2}$)\cite{moduli-mass}.
In fact if the moduli mass is $m_\phi \lesssim 200$ keV, 
the condition $\Omega_\phi \sim 1$ could be achieved
without conflict to the X($\gamma$)-ray observations
after an appropriate amount of late time entropy production is 
taken into account\cite{K-Y,H-K-Y,A-H-K-Y}.
Note that it  extensively dilutes various dark matter candidates
as well as the unwanted particles
and they could not give a sizable contribution to the energy density of
the universe.
Therefore, the string moduli seem to be a well motivated candidate 
for the CDM in the universe with the late time entropy production.%
\footnote{The cosmic axion with relatively high decay constant 
$f_{PQ} \sim 10^{16}$ GeV may be another candidate for the dark matter
\protect \cite{K-Y2,LSSS,K-M-Y}.}%

When we consider the moduli CDM,
we find that the saxion can play a role of the LDP
if the saxion is displaced at $\sigma_I \sim M_G$ after the 
primordial inflation.%
\footnote{
This initial condition is easily achieved by 
the effects of the supergravity.
}
Note that the moduli $\phi$ is also expected to be displaced from the true vacuum
at the order of the Planck scale, i.e., $\phi \sim M_G$.
The both fields begin to oscillate at almost the same time,
when the Hubble parameter becomes 
$H \simeq m_\phi \simeq m_\sigma (\simeq m_{3/2})$.
Using the initial amplitude of the moduli (saxion) oscillation
$\phi_0$ ($\sigma_0$),
the energy densities of the oscillations are related as
\begin{eqnarray}
    \label{Ratio-RsRp}
    \rho_\sigma 
    \simeq \rho_\phi
    \left( \frac{ \sigma_0}{\phi_0} \right)^2.
\end{eqnarray}
This equation also holds after some late time entropy production,
since the both energy densities are diluted at the same rate.
Therefore, the saxion abundance is given by 
\begin{eqnarray}
    \frac{\rho_\sigma}{s}
    &\simeq& \Omega_\phi \frac{\rho_c}{s_0}  
    \left( \frac{ \sigma_0}{\phi_0} \right)^2.
    \label{yield-s}
\end{eqnarray}
This abundance naturally explain the conditions 
of the saxion LDP Eqs.(\ref{cond-LDP1}) and (\ref{cond-LDP2}).
In fact, Eq.(\ref{cond-LDP2}) leads to the ratio of the initial amplitudes 
\begin{eqnarray}
    \frac{ \sigma_0 }{ \phi_0 } 
    &\simeq&
    \frac{ 0.32 }{ \Omega_\phi^{1/2} }
    \frac{ m_\sigma^{5/4} M_G^{1/4} }{ f \left( \rho_c /s_0
    \right)^{1/2} } \nonumber \\
    &\simeq&
    \frac{ 5.4 }{ \Omega_\phi^{1/2} }
    \left( \frac{ m_\sigma }{ 1~\mbox{MeV} } \right)^{5/4}
    \left( \frac{ f }{ 10^4~\mbox{GeV} } \right)^{-1},
\end{eqnarray}
with the saxion mass in the region (\ref{ms-LDP}).
Here the moduli CDM implies $\Omega_\phi \simeq 1$.
It should be noted that both initial amplitudes 
is expected to be of the order of the Planck scale,
$\sigma_0 \sim \phi_0 \sim M_G$.
Thus for $\Omega_\phi \simeq 1$ 
the saxion with the initial amplitude $\sigma_0$
which is slightly larger than $\phi_0 \sim M_G$
could be the LDP even with the late time entropy production
in the history of the universe,
if the saxion has a mass in the region (\ref{ms-LDP}).
On the other hand the string moduli
could be the CDM of our universe ($\Omega_\phi \simeq 1$)
if its mass is less than about 100 keV.
Therefore, if $m_\sigma \simeq m_\phi \simeq m_{3/2} \sim 100$ keV,
the string moduli could be the CDM of our universe while the saxion
could plays a role of the LDP so that the present observed 
power spectrum is well explained.

It is interesting that
this scenario will be tested in various cosmological experiments.
The moduli CDM with mass $m_\phi \sim $ 100 keV 
will be tested in the future by the x-ray background 
observations with high energy resolutions\cite{A-H-K-Y2}.
The line x-ray spectrum from the moduli CDM 
trapped in the halo of our galaxy will be seen. 
And the future CMBR observations, as mentioned before,
reveal the saxion LDP possibility.
%
\section{Conclusions}
%
In this paper, we closely investigated the cosmology of the
hadronic axion model in the gauge mediated supersymmetry breaking,
which we had proposed in a previous paper\cite{A-Y}. 
In particular we focused on
the implications given by the saxion which is the scalar component of
the axion supermultiplet. The saxion has a tiny mass comparable to the
gravitino mass and thus its life time tends to be long and decay after
the primordial nucleosynthesis. Furthermore since the saxion is a
scalar field, it will be abundant in the early universe in the form of
the coherent oscillation. Thus the saxion may be cosmologically harmful.

We estimated the energy density of the saxion in two
distinctive cases for the cosmological evolution of the saxion coherent
mode. Then we examined whether or not the saxion survives the various
cosmological constraints from the nucleosynthesis, from the critical
density limit, from the diffuse X-ray background spectrum, and from
the cosmic microwave background spectrum. We showed in both cases that
the model we proposed is cosmologically viable, namely it satisfies
the severe cosmological constraints on the saxion field, if the reheating
temperature of the primordial inflation is sufficiently low. 

Furthermore we argued that the saxion can play a role of the
late decaying particle in the structure formation of the universe, giving
a better fit of the power spectrum of the density perturbation than
the standard cold dark matter model.
\section*{acknowledgments}
We would like to thank M. Kawasaki and T. Yanagida for helpful discussions.
The work of MY was supported in part by the Grant-in-Aid for Scientific 
Research from the Ministry of Education, Science and Culture of Japan
No. 09640333.  

\clearpage
\end{document}